\documentclass[preprint,showpacs,aps,nofootinbib,superscriptaddress]{revtex4-1}
\usepackage{amssymb}
\usepackage{amsmath,amssymb,graphicx}
\usepackage{graphicx}
\usepackage{dcolumn}
\usepackage{bm}
\usepackage{multirow}

\def\beq{\begin{equation}}
\def\eeq{\end{equation}}
\def\bea{\begin{eqnarray}}
\def\eea{\end{eqnarray}}

\usepackage{pstricks}
\usepackage{graphicx}
\usepackage{amsbsy}
\usepackage{bm}
\usepackage{fancyhdr}
\usepackage{epsfig}
\usepackage[colorlinks=true,linkcolor=blue,citecolor=blue,urlcolor=blue,filecolor=blue]{hyperref}
\usepackage{slashed}

\begin{document}

\bigskip

\vspace{2cm}
\title{Lepton pair emission in the top quark decay $t\to bW^+ \ell^-\ell^+$ }  
\vskip 6ex
\author{N\'estor Quintero}
\email{nquintero@fis.cinvestav.mx}
\affiliation{Departamento de F\'isica, Centro de Investigaci\'on y de Estudios Avanzados, 
Apartado Postal 14-740, 07000 M\'exico D.F., M\'exico}
\author{J. Lorenzo D\'iaz-Cruz}
\email{jldiaz@fcfm.buap.mx}
\affiliation{Departamento de F\'isica, Centro de Investigaci\'on y de Estudios Avanzados, 
Apartado Postal 14-740, 07000 M\'exico D.F., M\'exico}
\affiliation{Facultad de Ciencias F\'isico Matem\'aticas, Universidad Aut\'onoma de Puebla, 
Apartado Postal 1152, Puebla, M\'exico}
\author{Gabriel L\'opez Castro}
\email{glopez@fis.cinvestav.mx}
\affiliation{Departamento de F\'isica, Centro de Investigaci\'on y de Estudios Avanzados, 
Apartado Postal 14-740, 07000 M\'exico D.F., M\'exico}

\bigskip
\begin{abstract}
The heaviness of the top quark makes its 2-body Cabibbo-favored Êdecay mode $t \to bW^+$ to be dominant, at such level that hardly any other decay mode reaches a detectable branching ratio (BR) within the SM. Here we study the decay $t \to bW^+\ell^-\ell^+$ ($\ell=e, \mu, \tau$), which diverges for massless leptons, and it can reach a BR $\sim \mathcal{O}(10^{-5}\sim 10^{-6})$ for reasonable values of the low energy cut in the lepton-pair invariant mass. This rate surpasses almost any other rare decays such as $t \to cX$ ($X=\gamma,Z,g,H, W^+W^-$), and thus offers the possibility of being detectable.  Furthermore, the estimate of this channel is relevant because it can mimic the 
signal arising from the lepton number violating  decay $t\to bW^-\ell^+\ell^+$, when 
the $W$ boson decays into lepton channels. 

\end{abstract}


\pacs{12.15.Ji, 14.60.-z14.65.Ha, 13.40.Ks}

\maketitle
\bigskip

\section{Introduction}

From our current explorations on the high energy frontier, we know that 
the top quark stands as the heaviest particle within the Standard Model (SM), which has been considered somehow peculiar and a hint that could help us 
to understand the nature of electroweak symmetry breaking. 
It is thus very important to study the top quark properties in order to search for these 
connections. In this regard, it is known that the top is also 
the only fermion massive enough to undergo first order  
weak decays, such that its dominant decay channel $t \to bW^+$ has 
a large decay width $\Gamma_t$, of similar size to the one for gauge bosons. 
This has encouraged the calculation of higher-order corrections to the rate of this 
decay channel: the first- \cite{first-qcd, denner} and second-order \cite{second-qcd} 
QCD corrections, the first order electroweak corrections \cite{denner,ew-corr} 
and the finite $W$ boson width effects \cite{first-qcd, finite-w} have been reported in 
the last twenty years. As it was summarized in ref. \cite{second-qcd}, 
these corrections turn out to be (in percent) $-8.58, -2.09, +1.69$ and $-1.49$ 
for $m_t=173.5$ GeV, respectively\footnote{In numerical evaluations we use this value of the top quark mass}. 

The dominance of the 2-body decay mode $t \to bW^+$ suppresses considerably any other decay channel,
making them hardly detectable. These include the decay modes $t \to s(d)W^+$, which contribute altogether 
less than one per mille to $\Gamma_t$.
In the case of 3-body decay modes, such as $t \to bW^+(\gamma,g)$ 
\cite{Tupper:1991,Couture:1989,Barger:1990,
Barger:1989,Mahlon:1998,Decker:1993}, $t \to bW^+Z$ 
\cite{Mahlon:1998,Decker:1993,Barger:1989,Altarelli:2001,Jenkins:1997,Mahlon:1995,Lorenzo:2000}, and 
$t \to bW^+H$ \cite{Barger:1989,Mahlon:1995,Mahlon:1998,Decker:1993,Rizzo:Barger,Han:2013}, the corresponding branching fractions 
(BR) are even smaller. Similarly, the FCNC modes $t \to cX$ ($X=\gamma,g,Z,H$) \cite{Mele:1998,LDCtopfv,Eilam:1991,Mele:1998,Aguilar:2003}, 
$t \to cW^+W^- (ZZ,\gamma\gamma)$ \cite{Jenkins:1997,Bar-Shalom:1998,Bar-Shalom:2005,Lorenzo:1999} and $t \to c\ell^-\ell^+$ \cite{Frank:2006}
have an extremely suppressed BR, although some of them involve very interesting 
dynamical mechanisms and have been suggested as possible probes of New Physics (NP), see for instance \cite{Aguilar:2004,Larios:2006}, 
which produces an enhancement on the BR that could make them detectable.  For the kinematically suppressed decay 
channels, the subsequent decay of unstable bosons is understood and sometimes 
have been taken into account. 

  In this paper we study the 4-body top quark 
decays $t \to b W^+ \ell^- \ell^+$  ($\ell =e, \mu$ or $\tau$). The 
rates of these decays are of order $\alpha^2$ with respect to $t\to bW$ and have a divergent behavior for 
massless leptons due to the $k^{-2}$ dependence of the photon propagator. For light leptons (see \cite{vanRitbergen:1998yd} for a related 
discussion in the case of the $\mu$ lifetime), this rate  should be included in the definition of the top quark width in 
order to  cancel the large  QED corrections of order $\mathcal{O}(\alpha^2)$ in $t \to bW^+$, arising from a light lepton 
loop-insertion in the infrared photon  propagator. Here, we are also interested in their study because they 
could mimic the signal from $t \to b W^-\ell^+\ell'^+$ when the  $W$ bosons are detected through 
its leptonic decays. The later is a lepton number ($\Delta L=2$) violating decay, 
which has been suggested as a signal of NP, induced by heavy Majorana 
neutrinos or doubly charged Higgs exchange \cite{top-lnv}. Some aspects 
related to the order $\alpha$ behavior of the radiative ($t \to bW^+\gamma$) and lepton-pair 
production ($t \to b W^+\ell^- \ell^+$) decay rates are discussed too. The top decay channel under consideration has not been widely studied before. To the best of our knowledge, only in reference \cite{Kong:2014jwa} it has been suggested that $t\to bW^+Z'$, with a subsequent decay of the $Z'$ boson into a lepton pair, may be a useful mechanism to detect a dark light $Z'$ gauge boson \cite{Essig:2013lka}.

The organization of this work goes as follows. In section \ref{radiativetop} we briefly recount the radiative 
decay $t \to bW^+\gamma$, in order to define notation and express the amplitude in such a way that is simple to 
identify the gauge invariance of the full amplitude. In section \ref{fourbody_top} we present the 4-body decay 
$t \to b W^+\ell^- \ell^+$, discussing in detail the dependence in calculation of the IR cutoff. 
Conclusions are left for section \ref{conclusions}.

\section{Radiative top quark decay $t \to b W^+\gamma$} \label{radiativetop}

\begin{figure}[!t]
\centering
\includegraphics[scale=0.6]{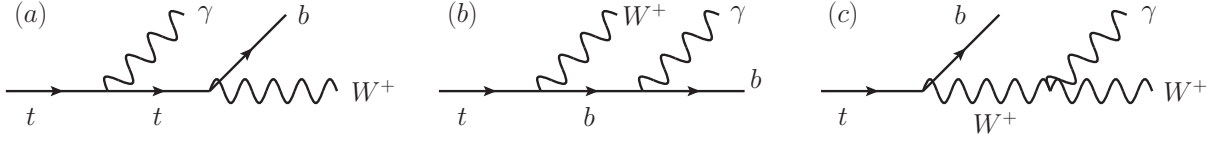}
\caption{\small Feynman diagrams that contribute to $t \to b W^+\gamma$.}
\label{Fig:topgamma}
\end{figure}

The radiative top quark  (3-body) decay $t(p_t) \to b(p_b) W^+(p_W)\gamma(k)$ has been 
widely considered in previous works \cite{Decker:1993,Tupper:1991,Couture:1989,Barger:1990,Barger:1989,Mahlon:1998}. 
We briefly discuss its amplitude, which is written in a form that is 
convenient to compare with the 4-body channel (sec. \ref{fourbody_top}); the 
numerical result for the decay rate is also included in order to discuss the infrared (IR) behavior. The Feynman diagrams that 
contribute in the unitary gauge are shown in Fig. \ref{Fig:topgamma}. Thus, the  
total decay amplitude  is written as follows
\bea \label{tbWgamma}
\mathcal{M}_{\text{top}}^{\gamma} &\equiv & \mathcal{M}(t \to b W^+\gamma), \nonumber \\
&=& \Big(\dfrac{-ig e}{\sqrt{2}} \Big)V_{tb} \ \varepsilon_\mu^W \varepsilon_\nu^\gamma \ \bar{u}(p_b) \Big[ Q_t \mathcal{T}_{t}^{\mu\nu}(k^2)  + Q_b \mathcal{T}_{b}^{\mu\nu}(k^2) + Q_W \mathcal{T}_{W}^{\mu\nu}(k^2) \Big] u(p_t),  
\eea

\noindent where 
\bea 
\mathcal{T}_{t}^{\mu\nu}(k^2) &=& \gamma^\mu P_L \dfrac{(\slashed{p_t} - 
\slashed{k})  + m_t}{k^2 - 2  p_t\cdot k} \gamma^\nu, \label{T1} \\
\mathcal{T}_{b}^{\mu\nu}(k^2) &=& \gamma^\nu \dfrac{(\slashed{p_b} + 
\slashed{k}) + m_b}{k^2 + 2  p_b\cdot k} \gamma^\mu P_L , \label{T2} \\
\mathcal{T}_{W}^{\mu\nu}(k^2) &=& \gamma^\alpha P_L \Delta^{W}_{\alpha\beta}(h) 
\Gamma_{WW\gamma}^{\beta\mu\nu}. \label{T3} 
\eea 

\noindent The amplitude for a real photon emission is obtained by taking $k^2 =0$ in the above expressions. $P_L $ denotes the left-handed chiral 
projector, ($Q_b , Q_t, Q_W$) are the particles electric charges in 
units of $e$, and $\varepsilon^W_{\mu}$ ($\varepsilon_\nu^\gamma$) is 
the four-vector polarization of the $W$ boson (photon), respectively. 
The CKM quark mixing matrix element is taken as $V_{tb} = 1$. The propagator 
of the $W$ boson in  the unitary gauge is denoted by $\Delta^W_{\alpha\beta}(h) = (-g_{\alpha\beta}+h_{\alpha}h_{\beta}/M_W^2)/(h^2 - M_W^2)$, 
with $h=p_W+k$, while the triple gauge boson vertex, with our assignment of momenta 
$W(p_W+k)\to W(p_W)\gamma(k)$, is given by
\beq
\Gamma^{\beta\mu\nu}_{WW\gamma} = (p_W - k)^\beta g^{\mu\nu} - 
(2p_W + k)^\nu g^{\mu\beta} + (2k + p_W)^\mu g^{\nu\beta}.
\eeq 

The decay amplitude $\mathcal{M}_{\text{top}}^{\gamma}$ is gauge-invariant 
owing to the charge conservation condition $Q_b - Q_t + Q_W = 0$ \cite{Couture:1989} 
and diverges in the soft-photon limit ($k \to 0$). The photon energy spectrum is obtained by 
integrating the unpolarized squared amplitude over $s_1=(p_b + k)^2$, namely: 
\bea
\frac{d\Gamma(t \to b W^+\gamma)}{dE_{\gamma}} = \dfrac{1}{128 \pi^3 m_t^2}  
\int_{s_{1}^-}^{s_{1}^+} ds_{1} \ |\mathcal{M}_{\text{top}}^{\gamma}|^2.
\eea
For completeness we show here the integration limits:
\bea
s_{1}^{\pm}(s_2) &=& M_{W}^{2} + \dfrac{m_t^2-s_2}{2 s_2} \Big[(s_2 - m_{b}^{2} + M_{W}^2) \pm \sqrt{\lambda(s_2,M_{W}^{2},m_b^{2})} \Big],
\eea
with $s_2 =m_t^2 - 2m_t E_\gamma$ and $\lambda(x,y,z)=x^2+y^2+z^2-2(xy+xz+yz)$. 
The photon energy goes from $0$ up to a maximun value $E_{\gamma}^{\text{max}} = [m_t^2 - (M_W + m_b)^2]/2m_t$. 
However, because of the IR divergence, it is necessary to impose a cut in the minimun photon energy ($E_{\gamma,\text{cut}}$) 
in order to obtain a finite value.

\begin{figure}[!t]
\centering
\includegraphics[scale=0.5]{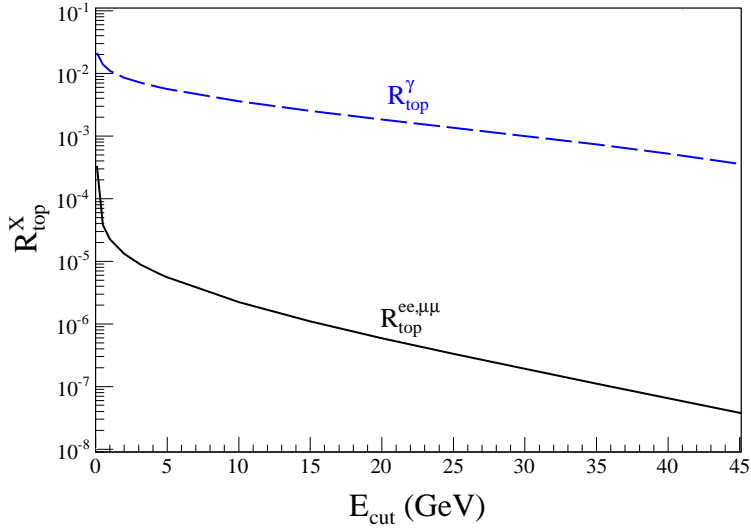}\\
\caption{\small Ratios  R$_\text{top}^X$ as a function of $E_\text{cut}$ for the radiative channel ($X=\gamma$) and lepton-pair channels ($X=e^-e^+,\mu^-\mu^+$).}
\label{Comparison}
\end{figure}

The normalized radiative rate is defined as R$^\gamma_{\text{top}} \equiv \Gamma(t \to b W^+\gamma)/ 
\Gamma(t \to b W^+)$, where the 2-body decay width at leading order is given by  \cite{PDG}
\beq
\Gamma(t \to b W^+) = \dfrac{G_F m_t^3}{8\pi\sqrt{2}} (1- x_t^2)^2 (1+2 x_t^2),
\eeq
with $x_t = M_W/m_t$. R$^\gamma_{\text{top}}$ is plotted in Fig. \ref{Comparison} (dashed line) as a function of 
$E_{\gamma,\text{cut}}$. Our results are consistent with those obtained 
in Refs. \cite{Decker:1993,Tupper:1991,Couture:1989,Barger:1990,Barger:1989,
Mahlon:1998}. It is well known that the soft-photon divergence in R$^\gamma_{\text{top}}$ will 
be cancelled by the corresponding infrared divergence from the one-loop virtual photon 
corrections, giving rise to a finite photon inclusive rate $\Gamma(t \to bW^+(\gamma))$. 
As it can be seen from Fig. \ref{Comparison}, the expected $\mathcal{O}(\alpha)$ suppression of the radiative decay, with respect to the dominant 2-body decay mode, is already evident for $E_{\gamma,\text{cut}} \simeq 3$ GeV.

\section{Four-body decay $t \to b W^+\ell^-\ell^+$} \label{fourbody_top}

\begin{figure}[!t]
\centering
\includegraphics[scale=0.55]{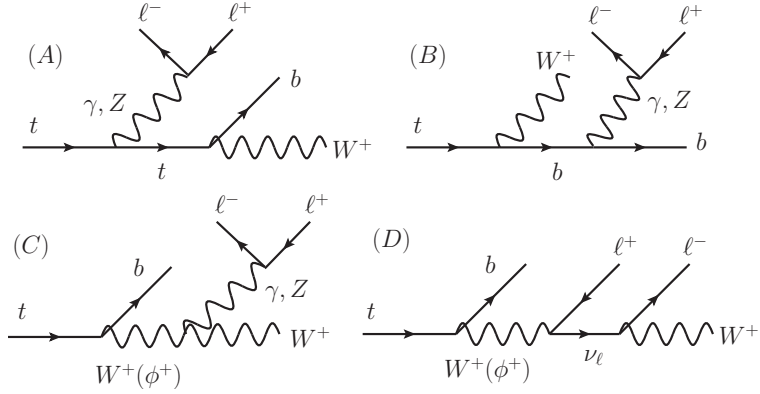}\
\caption{\small Feynman diagrams that contribute to $t \to b W^+ \ell^-\ell^+$.}
\label{Fig:Leptonpair}
\end{figure}

Now, let us consider the lepton-pair production in top quark decays  
$t(p_t) \to b(p_b) W^+(p_W)$ $\ell^-(p_1)\ell^+(p_2)$ with $\ell =e,\mu$ or $\tau$. 
The relevant diagrams for this decay are shown in Fig. \ref{Fig:Leptonpair}.  
The diagrams  Fig. \ref{Fig:Leptonpair}($A$)-($C$) are generated by demanding 
that the virtual photon ($Z$ boson) converts into a lepton-pair\footnote{The 
lepton-pair can also be produced by an intermediate Higgs boson, however this gives
a negligible contribution.}. An  additional contribution induced by a neutrino exchange 
is shown in Fig. \ref{Fig:Leptonpair}($D$). The full set of diagrams is required 
in order to fulfill independence upon the electroweak gauge parameters.

The dominant photon-exchange contribution to the amplitude is written as  
\bea \label{tbWll}
\mathcal{M}_{\text{top}}^{\ell\ell} &\equiv & \mathcal{M}
(t \to b W^+\ell^-\ell^+), \nonumber \\
 &=& \Big(\dfrac{-ig e}{\sqrt{2}} \Big)V_{tb} \ \varepsilon_\mu^W \ell_\nu 
\ \bar{u}(p_b) \Big[ Q_t \mathcal{T}_{t}^{\mu\nu}(k^2)  + Q_b \mathcal{T}_{b}^{\mu\nu}(k^2) 
+ Q_W \mathcal{T}_{W}^{\mu\nu}(k^2) \Big] u(p_t),  
\eea

\noindent where $\ell_\nu = e[\bar{u}(p_1) \gamma_\nu v(p_2)] / k^2$, 
with $k =p_1 + p_2$ the virtual photon momentum. The $\mathcal{T}_i^{\mu\nu}(k^2)$ 
tensors are given in Eqs. (\ref{T1})-(\ref{T3}), this time keeping $k^2 \not =0$ 
in the propagators.  In fact, the amplitude $\mathcal{M}_{\text{top}}^{\ell\ell}$, 
Eq. (\ref{tbWll}), is easily obtained from Eq. (\ref{tbWgamma}) by replacing the 
photon polarization $(\varepsilon_\nu^\gamma)$ by the effective leptonic current $(\ell_\nu)$. 

The contributions of the $Z$-boson exchange amplitude (Fig. \ref{Fig:Leptonpair}A-C) are suppressed owing to the large mass of
 this gauge boson. An estimate of its largest contribution can be obtained when the $Z$ boson is taken on-shell. In this case the branching ratio for 
$t \to b W^+ \ell^-\ell^+$ can be estimated from the SM prediction for 
$\Gamma(t \to bW^+Z) / \Gamma(t \to bW^+) \simeq 2\times 10^{-6}$ 
\cite{Altarelli:2001} followed by the leptonic decay of the $Z$ boson, which yields:
\beq
\dfrac{\Gamma(t \to bW^+Z)\times \mathcal{B}(Z \to \ell^-\ell^+)}{\Gamma(t \to bW^+)} 
 \simeq 6\times 10^{-8}.
\eeq

\noindent As it will be shown, this is smaller than the photon contribution.
Finally, the neutrino exchange diagram (Fig. \ref{Fig:Leptonpair}($D$)) is necessary to 
guarantee a gauge-invariant result for the full decay amplitude \cite{Altarelli:2001}, 
yet its contribution is even smaller that the one from $Z$ boson exchange.
Thus, the leading contribution is given by Eq. (\ref{tbWll}). 

\begin{figure}[!t]
\centering
\includegraphics[scale=0.45]{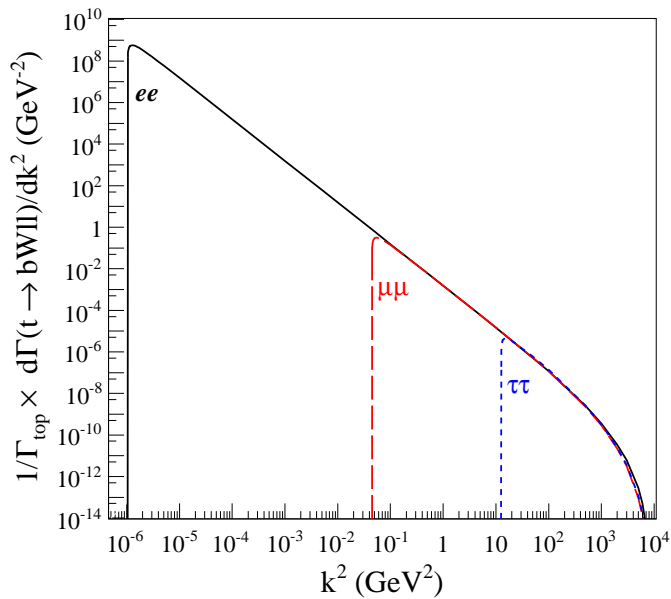}\\
\caption{\small Normalized invariant-mass distribution for 
$e^-e^+$ (solid line), $\mu^-\mu^+$ (long-dashed line) and $\tau^-\tau^+$ (short-dashed line) channels.}
\label{Spectrum}
\end{figure}

In Figure \ref{Spectrum} we plot the invariant mass distribution of the lepton pair 
for the three lepton flavors. This observable is peaked close to the threshold for 
lepton pair production owing to the $1/k^2$ dependence of the photon propagator from 
the decay amplitude. As it can be checked, this spectrum diverges in the soft 
 limit for massless leptons ($k^2 \to 0$); this divergence would 
be eventually cancelled by the corresponding IR divergence resulting from the QED corrections to $t \to bW^+$ decay at two-loops, specifically from the massless lepton loop insertion to the photon propagators. In practice, since electrons and muons have a small (but nonzero) mass, the branching ratio may become meaninglessly large when integrating over the full range of $k^2$, making necessary the introduction of an infrared cutoff $k^2_{\rm cut}$.

Given the very steeply behavior of the squared amplitude close to the lepton pair 
threshold,  it is convenient to check the stability of our numerical result. This is done by evaluating the  ratio R$_{\rm top}^{\ell\ell}\equiv \Gamma(t\to bW^+\ell^-\ell^+)/\Gamma(t\to bW^+)$ (with $\ell=e, \mu$, $\tau$) using two numerical methods: $(i)$ integrating over the 
lepton-pair invariant mass distribution and,  $(ii)$ by integrating upon 
the five independent kinematical variables of the four-body phase space. We have
checked that both methods give identical results, except when the integration over $k^2$ is extended until the lepton-pair threshold in the case of electrons.   The R$_{\rm top}^{ee, \mu\mu}$ ratios 
   as a function of $k^2_\text{cut}$ are shown 
in Table \ref{TableI}. 
As previously advertised,  the  ratio  R$_{\rm top}^{ee}$ for producing 
an $e^-e^+$ pair becomes larger than unity for 
$k^2_\text{cut} < 10^{-3}$ GeV$^2$. Although for lepton pairs $\mu^-\mu^+$ and $\tau^-\tau^+$  the obtained value is not greater than one, the result in the case of muons is still unexpectedly large from the perturbative point of view. However, as we can see from Table \ref{TableI}, in all cases the expected behavior for the branching ratio,  of $O(10^{-5}\sim 10^{-6})$ 
as determined by the $\mathcal{O}(\alpha^2)$ of the decay rates, can be 
obtained for values of $k^2_\text{cut} \geq 20$  GeV$^2$.

\begin{table}[!t]
\label{explimits}
\renewcommand{\arraystretch}{1.2}
\renewcommand{\arrayrulewidth}{0.8pt} 
\caption{The ratio R$_\text{top}^{\ell\ell}$ ($\ell = e, \mu$, $\tau$) as a function of $k^2_\text{cut}$. The instability for the result in the first entry of  R$_\text{top}^{ee}$ is indicated with a star symbol }
\begin{tabular}{cccc}
\hline\hline 
$k^2_\text{cut}$ (GeV$^2$) & R$_\text{top}^{ee}$  & R$_\text{top}^{\mu\mu}$ & R$_\text{top}^{\tau\tau}$\\
 \hline
$4m_\ell^2$ & 6.49$^\star$ & $1.52\times 10^{-2}$ & $1.06 \times 10^{-5}$ \\
$10^{-3}$ & 0.85 & -- & -- \\
$10^{-2}$ & $8.52\times 10^{-2}$ & -- & --  \\
$10^{-1}$ & $8.48\times 10^{-3}$ & $8.48\times 10^{-3}$ & -- \\
$1$ & $8.28\times 10^{-4}$ & $8.28\times 10^{-4}$ & --\\
$10$ & $7.40\times 10^{-5}$ & $7.40\times 10^{-5}$ & -- \\
$20$ & $6.31\times 10^{-6}$ & $6.31\times 10^{-6}$ & $9.15 \times 10^{-6}$ \\
$50$ & $3.70\times 10^{-6}$ & $3.70\times 10^{-6}$ & $5.50 \times 10^{-6}$ \\
\hline\hline
\end{tabular} \label{TableI}
\end{table}

In Figure \ref{Comparison}, we compare the IR divergent behavior of the normalized rates for  
$t \to bW^+e^-e^+ (\mu^-\mu^+)$ (solid-line) and $t \to bW^+\gamma$ (dashed-line) as a function of $E_\text{cut}$. By notation, in the case of the radiative decay $E_\text{cut} = E_{\gamma,\text{cut}}$ corresponds to the photon energy, while for the four-body channel $E_\text{cut} = \sqrt{k^2_\text{cut}}$ is the squared root of the invariant mass for the lepton pair. 
We observe that these ratios 
are of order $\mathcal{O}(\alpha^2)$ and $\mathcal{O}(\alpha)$, respectively, 
for relatively low values of the corresponding IR cutoffs (of the order $E_\text{cut} \simeq 3$ GeV). Above this 
critical value, the ratio between these two decay modes becomes of order $\mathcal{O}(\alpha)$ as it 
would be expected.


\section{Conclusions} \label{conclusions}

After the success of the first stage of the LHC, with the detection of the Higgs boson as its 
shining trophy, we hope that the next one, with higher energy, will continue to test the SM and hopefully find
a signal of new physics. In particular, the LHC will also become a productive top factory, which will be able
to test the top properties, its couplings to SM channels and rare decays. Having about $10^7-10^8$ top pairs 
produced per year at LHC, it is expected that rare decays with BR of order $10^{-5}-10^{-6}$ may be
detectable, depending on the signal.

Along these lines, we have presented in this paper the (first) calculation of the process where a lepton pair 
is emited in top quark decays. Both, the spectrum and the branching ratios 
exhibit an infrared divergent  behaviour in the limit of massless leptons. We observed 
that by cutting at $\ell^-\ell^+$ invariant masses larger than 20 GeV$^2$, 
the branching ratios follow the suppression rule expected by their relative order in $\alpha^2$. 

The lepton pair production processes $t \to bW^+ (\to \ell'^+ \nu_{\ell}) \ell^+\ell^-$  can provide and important  background 
for searches of  lepton number violating (LNV) top quark decays  $t\to bW^-(\to \ell^- \bar{\nu})\ell^+\ell'^+$, when $W$ bosons are 
detected through lepton channels. Table \ref{TableI} shows that even by cutting the large phase space in the lepton-pair invariant mass, the fraction of  
conventional decays is larger than the LNV decays, whose most optimistic estimates are of order a few $\times 10^{-6}$ \cite{top-lnv}. 

\medskip

{\bf Acknowledgements}: The authors would like to thank Conacyt (M\'exico) 
for financial support. They are also grateful to A. Guevara and E. Camacho 
for useful conversations.


\end{document}